\documentstyle[12pt]{article}
\newcommand{\nc}{\newcommand}
\nc{\renc}{\renewcommand}

\nc{\etal}{\mbox{\it et al. }}
\nc{\ie}{{\it i.e.}}
\nc{\eg}{{\it e.g.}}

\renc{\thefootnote}{\arabic{footnote}}
\nc{\capt}[1]{{\bf Figure.} {\small\sl #1}}

\nc{\eqs}[2]{\mbox{Eqs.~(\ref{#1},\,\ref{#2})}}
\nc{\eq}[1]{\mbox{Eq.~(\ref{#1})}}

\nc{\figs}[2]{\mbox{Figs.~(\ref{#1},\,\ref{#2})}}
\nc{\fig}[1]{\mbox{Fig~.(\ref{#1})}}

\nc{\tag}[1]{\label{#1} \marginpar{{\footnotesize #1}}}
\nc{\mtag}[1]{\label{#1} \mbox{\marginpar{{\footnotesize #1}}}}

\renc{\baselinestretch}{1.2}
\newlength{\overeqskip}
\newlength{\undereqskip}
\nc{\be}[1]{\begin{equation} \mbox{$\label{#1}$}}
\nc{\bea}[1]{\begin{eqnarray} \mbox{$\label{#1}$}}
\nc{\Section}[2]{\section{#2}\label{#1}}
\nc{\Bibitem}[1]{\bibitem{#1}}
\nc{\Label}[1]{\label{#1}}

\nc{\eea}{\vspace{\undereqskip}\end{eqnarray}}
\nc{\ee}{\vspace{\undereqskip}\end{equation}}
\nc{\bdm}{\begin{displaymath}}
\nc{\edm}{\end{displaymath}}
\nc{\dpsty}{\displaystyle}
\nc{\bc}{\begin{center}}
\nc{\ec}{\end{center}}
\nc{\ba}{\begin{array}}
\nc{\ea}{\end{array}}
\nc{\bab}{\begin{abstract}}
\nc{\eab}{\end{abstract}}
\nc{\btab}{\begin{tabular}}
\nc{\etab}{\end{tabular}}
\nc{\bit}{\begin{itemize}}
\nc{\eit}{\end{itemize}}
\nc{\ben}{\begin{enumerate}}
\nc{\een}{\end{enumerate}}
\nc{\bfig}{\begin{figure}}
\nc{\efig}{\end{figure}}
\nc{\arreq}{&\!=\!&}
\nc{\arrmi}{&\!-\!&}
\nc{\arrpl}{&\!+\!&}
\nc{\arrap}{&\!\!\!\approx\!\!\!&}
\nc{\non}{\nonumber\\*}
\nc{\align}{\!\!\!\!\!\!\!\!&&}

\def\lsim{\; \raise0.3ex\hbox{$<$\kern-0.75em
      \raise-1.1ex\hbox{$\sim$}}\; }
\def\gsim{\; \raise0.3ex\hbox{$>$\kern-0.75em
      \raise-1.1ex\hbox{$\sim$}}\; }
\nc{\DOT}{\hspace{-0.08in}{\bf .}\hspace{0.1in}}
\nc{\Laada}{\hbox {$\sqcap$ \kern -1em $\sqcup$}}
\nc\loota{{\scriptstyle\sqcap\kern-0.55em\hbox{$\scriptstyle\sqcup$}}}
\nc\Loota{{\sqcap\kern-0.65em\hbox{$\sqcup$}}}
\nc\laada{\Loota}
\nc{\qed}{\hskip 3em \hbox{\BOX} \vskip 2ex}
\def\Re{{\rm Re}\hskip2pt}
\def\Im{{\rm Im}\hskip2pt}
\nc{\real}{{\rm I \! R}}
\nc{\Z}{{\sf Z \!\!\! Z}}
\nc{\complex}{{\rm C\!\!\! {\sf I}\,\,}}
\def\bigid{\leavevmode\hbox{\small1\kern-3.8pt\normalsize1}}
\def\id{\leavevmode\hbox{\small1\kern-3.3pt\normalsize1}}
\nc{\slask}{\!\!\!/}
\nc{\bis}{{\prime\prime}}
\nc{\pa}{\partial}
\nc{\na}{\nabla}
\nc{\ra}{\rangle}
\nc{\la}{\langle}
\nc{\goto}{\rightarrow}
\nc{\swap}{\leftrightarrow}

\nc{\EE}[1]{ \mbox{$\cdot10^{#1}$} }
\nc{\abs}[1]{\left|#1\right|}
\nc{\at}[2]{\left.#1\right|_{#2}}
\nc{\norm}[1]{\|#1\|}
\nc{\abscut}[2]{\Abs{#1}_{\scriptscriptstyle#2}}
\nc{\vek}[1]{{\rm\bf #1}}
\nc{\integral}[2]{\int\limits_{#1}^{#2}}
\nc{\inv}[1]{\frac{1}{#1}}
\nc{\dd}[2]{{{\partial #1}\over{\partial #2}}}
\nc{\ddd}[2]{{{{\partial}^2 #1}\over{\partial {#2}^2}}}
\nc{\dddd}[3]{{{{\partial}^2 #1}\over
	{\partial #2 \partial #3}}}
\nc{\dder}[2]{{{d #1}\over{d #2}}}
\nc{\ddder}[2]{{{d^2 #1}\over{d {#2}^2}}}
\nc{\dddder}[3]{{d^2 #1}\over
	{d #2 d #3}}
\nc{\dx}[1]{d\,^{#1}x}
\nc{\dy}[1]{d\,^{#1}y}
\nc{\dz}[1]{d\,^{#1}z}
\nc{\dl}[1]{\frac{d\,^{#1}l}{(2\pi)^{#1}}}
\nc{\dk}[1]{\frac{d\,^{#1}k}{(2\pi)^{#1}}}
\nc{\dq}[1]{\frac{d\,^{#1}q}{(2\pi)^{#1}}}

\nc{\cc}{\mbox{$c.c.$ }}
\nc{\hc}{\mbox{$h.c.$ }}
\nc{\cf}{cf.\ }
\nc{\erfc}{{\rm erfc}}
\nc{\Tr}{{\rm Tr\,}}
\nc{\tr}{{\rm tr\,}}
\nc{\pol}{{\rm pol}}
\nc{\sign}{{\rm sign}}
\nc{\bfT}{{\bf T }}

\nc{\cA}{{\cal A}}
\nc{\cB}{{\cal B}}
\nc{\cD}{{\cal D}}
\nc{\cE}{{\cal E}}
\nc{\cG}{{\cal G}}
\nc{\cH}{{\cal H}}
\nc{\cL}{{\cal L}}
\nc{\cO}{{\cal O}}
\nc{\cT}{{\cal T}}
\nc{\cN}{{\cal N}}
\nc{\rvac}[1]{|{\cal O}#1\rangle}
\nc{\lvac}[1]{\langle{\cal O}#1|}
\nc{\rvacb}[1]{|{\cal O}_\beta #1\rangle}
\nc{\lvacb}[1]{\langle{\cal O}_\beta #1 |}
\nc{\bb}{\bar{\beta}}
\nc{\bt}{\tilde{\beta}}
\nc{\ctH}{\tilde{\cal H}}
\nc{\chH}{\hat{\cal H}}
\nc{\1}{\aa}
\nc{\2}{\"{a}}
\nc{\3}{\"{o}}
\nc{\4}{\AA}
\nc{\5}{\"{A}}
\nc{\6}{\"{O}}
\nc{\al}{\alpha}
\nc{\g}{\gamma}
\nc{\Del}{\Delta}
\nc{\e}{\epsilon}
\nc{\eps}{\epsilon}
\nc{\lam}{\lambda}
\nc{\om}{\omega}
\nc{\Om}{\Omega}
\nc{\ve}{\varepsilon}
\nc{\mn}{{\mu\nu}}
\nc{\k}{\kappa}
\nc{\vp}{\varphi}

\nc{\advp}[3]{{\it  Adv.\ in\ Phys.\ }{{\bf #1} {(#2)} {#3}}}
\nc{\annp}[3]{{\it  Ann.\ Phys.\ (N.Y.)\ }{{\bf #1} {(#2)} {#3}}}
\nc{\apl}[3]{{\it  Appl. Phys. Lett. }{{\bf #1} {(#2)} {#3}}}
\nc{\apj}[3]{{\it  Ap.\ J.\ }{{\bf #1} {(#2)} {#3}}}
\nc{\apjl}[3]{{\it  Ap.\ J.\ Lett.\ }{{\bf #1} {(#2)} {#3}}}
\nc{\app}[3]{{\it Astropart.\ Phys.\ }{{\bf #1} {(#2)} {#3}}}  
\nc{\cmp}[3]{{\it  Comm.\ Math.\ Phys.\ }{{ \bf #1} {(#2)} {#3}}}
\nc{\cqg}[3]{{\it  Class.\ Quant.\ Grav.\ }{{\bf #1} {(#2)} {#3}}}
\nc{\epl}[3]{{\it  Europhys.\ Lett.\ }{{\bf #1} {(#2)} {#3}}}
\nc{\ijmp}[3]{{\it Int.\ J.\ Mod.\ Phys.\ }{{\bf #1} {(#2)} {#3}}}
\nc{\ijtp}[3]{{\it Int.\ J.\ Theor.\ Phys.\ }{{\bf #1} {(#2)} {#3}}}
\nc{\jmp}[3]{{\it  J.\ Math.\ Phys.\ }{{ \bf #1} {(#2)} {#3}}}
\nc{\jpa}[3]{{\it  J.\ Phys.\ A\ }{{\bf #1} {(#2)} {#3}}}
\nc{\jpc}[3]{{\it  J.\ Phys.\ C\ }{{\bf #1} {(#2)} {#3}}}
\nc{\jap}[3]{{\it J.\ Appl.\ Phys.\ }{{\bf #1} {(#2)} {#3}}}
\nc{\jpsj}[3]{{\it J.\ Phys.\ Soc.\ Japan\ }{{\bf #1} {(#2)} {#3}}}
\nc{\lmp}[3]{{\it Lett.\ Math.\ Phys.\ }{{\bf #1} {(#2)} {#3}}}
\nc{\mpl}[3]{{\it  Mod.\ Phys.\ Lett.\ }{{\bf #1} {(#2)} {#3}}}
\nc{\ncim}[3]{{\it  Nuov.\ Cim.\ }{{\bf #1} {(#2)} {#3}}}
\nc{\np}[3]{{\it  Nucl.\ Phys.\ }{{\bf #1} {(#2)} {#3}}}
\nc{\pr}[3]{{\it Phys.\ Rev.\ }{{\bf #1} {(#2)} {#3}}}
\nc{\pra}[3]{{\it  Phys.\ Rev.\ A\ }{{\bf #1} {(#2)} {#3}}}
\nc{\prb}[3]{{\it  Phys.\ Rev.\ B\ }{{{\bf #1} {(#2)} {#3}}}}
\nc{\prc}[3]{{\it  Phys.\ Rev.\ C\ }{{\bf #1} {(#2)} {#3}}}
\nc{\prd}[3]{{\it  Phys.\ Rev.\ D\ }{{\bf #1} {(#2)} {#3}}}
\nc{\prl}[3]{{\it Phys\ Rev.\ Lett.\ }{{\bf #1} {(#2)} {#3}}}
\nc{\pl}[3]{{\it  Phys.\ Lett.\ }{{\bf #1} {(#2)} {#3}}}
\nc{\prep}[3]{{\it Phys.\ Rep.\ }{{\bf #1} {(#2)} {#3}}}
\nc{\prsl}[3]{{\it Proc.\ R.\ Soc.\ London\ }{{\bf #1} {(#2)} {#3}}}
\nc{\ptp}[3]{{\it  Prog.\ Theor.\ Phys.\ }{{\bf #1} {(#2)} {#3}}}
\nc{\ptps}[3]{{\it  Prog\ Theor.\ Phys.\ suppl.\ }{{\bf #1} {(#2)} {#3}}}
\nc{\physa}[3]{{\it  Physica\ A\ }{{\bf #1} {(#2)} {#3}}}
\nc{\physb}[3]{{\it  Physica\ B\ }{{\bf #1} {(#2)} {#3}}}
\nc{\phys}[3]{{\it Physica\ }{{\bf #1} {(#2)} {#3}}}
\nc{\rmp}[3]{{\it  Rev.\ Mod.\ Phys.\ }{{\bf #1} {(#2)} {#3}}}
\nc{\rpp}[3]{{\it Rep.\ Prog.\ Phys.\ }{{\bf #1} {(#2)} {#3}}}
\nc{\sjnp}[3]{{\it Sov.\ J.\ Nucl.\ Phys.\ }{{\bf #1} {(#2)} {#3}}}
\nc{\spjetp}[3]{{\it Sov.\ Phys.\ JETP\ }{{\bf #1} {(#2)} {#3}}}
\nc{\yf}[3]{{\it Yad.\ Fiz.\ }{{\bf #1} {(#2)} {#3}}}
\nc{\zetp}[3]{{\it Zh.\ Eksp.\ Teor.\ Fiz.\  }{{\bf #1}  {(#2)} {#3}}}
\nc{\zp}[3]{{\it Z.\ Phys.\ }{{\bf #1} {(#2)} {#3}}}
\nc{\ibid}[3]{{\sl ibid.\ }{{\bf #1} {#2} {#3}}}
\nc{\rf}[1]{(\ref{#1})}
\nc{\nn}{\nonumber \\*}
\nc{\bfB}{\bf{B}}
\nc{\bfv}{\bf{v}}
\nc{\bfx}{\bf{x}}
\nc{\bfy}{\bf{y}}
\nc{\vx}{\vec{x}}
\nc{\vy}{\vec{y}}
\nc{\oB}{\overline{B}}
\nc{\oI}{\overline{I}}
\nc{\oR}{\overline{R}}
\nc{\rar}{\rightarrow}
\nc{\ti}{\times}
\nc{\slsh}{\hskip-5pt/}
\nc{\sm}{Standard~Model~}
\nc{\MP}{M_{\rm Pl}}
\nc{\tp}{t_{\rm Pl}}
\nc{\ave}{\bar{E}}

\renc{\min}{p_{\rm min}}
\renc{\max}{p_{\rm max}}
\nc{\pmin}{p_{\rm min}}
\nc{\pmax}{p_{\rm max}}
\nc{\fo}{f_0}
\nc{\foi}{f_{0,i}\,}
\nc{\fop}{f_0^P}
\nc{\fou}{f_0^U}
\def\sepand{\rule{14cm}{0pt}\and}
\nc{\eff}{{\rm eff}}
\nc{\MT}{M_{\rm T}}
\nc{\ML}{M_{\rm L}}
\nc{\kk}{\vek{k}}
\nc{\pp}{{\rm p}}
\nc{\cb}{critical bubble~}
\nc{\cbs}{critical bubbles~}
\nc{\scb}{subcritical bubble~}
\nc{\scbs}{subcritical bubbles~}
\nc{\MSSM}{Minimal Supersymmetric Standard Model}
\nc{\mato}[1]{\mathaccent 126 #1}
\nc{\vii}[1]{(\ref #1)}

\begin{document}

{\title{{\hfill {{\small  TURKU-FL-P28-98 
        }}\vskip 1truecm}
{\bf Spontaneous R-Parity Violation and Electroweak Baryogenesis}}

 
\author{
{\sc Tuomas Multam\" aki$^{1}$}\\
{\sl and}\\
{\sc Iiro Vilja$^{2}$ }\\ 
{\sl Department of Physics,
University of Turku} \\
{\sl FIN-20500 Turku, Finland} \\\sepand
}
\date{April 23, 1998}
\maketitle}
\vspace{2cm}
\begin{abstract}
\noindent 
The possibility of baryogenesis at the electroweak phase transition
is considered within
the context of a minimal supersymmetric standard model with spontaneous
R-parity violation. 
Provided that at least one of the sneutrino fields acquires a large enough
vacuum expectation value, a sufficient baryon asymmetry can be created.
Compared to R-parity conserving models the choice of soft supersymmetry 
breaking parameters is less restricted. The observed baryon 
asymmetry, ${n_B\over s}\sim10^{-10}$, can be explained by this scenario and 
the produced baryon-to-entropy ratio may easily be as high as 
${n_B\over s}\sim 10^{-9}$. \end{abstract}
\vfill
\footnoterule
{\small$^1$tuomul@newton.tfy.utu.fi,  $^2$vilja@newton.tfy.utu.fi}
\thispagestyle{empty}
\newpage
\setcounter{page}{1}
One of the major open questions in theoretical particle physics is 
to explain the observed baryon asymmetry of the universe. 
The well-known Sakharov conditions for baryogenesis, namely
T, B, C and CP -violation \cite{sakh}, can be fulfilled in
the early universe during a phase transition for suitable
models of particle physics.
In particular, it has been 
shown that the Standard Model (SM) includes all the requirements for 
electroweak baryogenesis \cite{rev}.
However, the electroweak phase transition is too
weakly first order to preserve the generated baryon asymmetry in the
SM \cite{shapo}. Furthermore, the CP-violation due to the phase of the 
Cobayashi-Maskawa -matrix is too small for a sufficient generation 
of baryons \cite{gavela}. The realization of 
electroweak baryogenesis therefore requires physics beyond the SM.

The \MSSM  (MSSM) has appeared to be a \\promising 
candidate for the explanation of the origin of
the observed baryon asymmetry in the universe.
It has been shown that with an appropriate choice of parameters
a large enough baryon asymmetry, ${n_B\over s}\sim 10^{-10}$, could have 
been created 
at the electroweak phase transition \cite{strength1,CQRVW}.
The necessary choice of parameters involves, however, 
quite stringent bounds. Furthermore the choice of $\abs{\sin\phi_\mu}$,
where $\phi_\mu$ is the phase of one of the soft supersymmetry breaking
parameters $\mu$,
has been constrained even more in a more recent analysis of the
bubble wall profile \cite{multamaki}. It is therefore useful and
interesting to expand the idea of electroweak baryogenesis to 
less constrained supersymmetric models. In this paper
we shall consider the creation of baryon asymmetry in the context
of a model where R-parity is spontaneously violated \cite{rviolation}.

The \MSSM\ (MSSM) assumes the conservation of a discrete symmetry,
R-parity \cite{haberkane,nilles}. R-parity is related to the spin (S),
total lepton (L) and baryon (B) number of a particle by
$R_p=(-1)^{3B+L+2S}$ so that all the standard model particles have an
even R-parity whereas their supersymmetric partners have an odd
R-parity. There is, however, no {\it a priori} reason to expect that
R-parity is conserved. Therefore it can be a spontaneously or
even explicitly broken symmetry.
A model that spontaneously violates R-parity 
proposed in 
ref. \cite{masiero} has a superpotential of the form (suppressing
all $SU(2)$ and generation indices)
\be{superp}
h_u\hat{Q}\hat{H}_u\hat{u}^c+h_d\hat{H}_d\hat{Q}\hat{d}^c+h_e\hat{l}
\hat{H}_d\hat{e}^c+(h_0\hat{H}_u\hat{H}_d-\mu^2)\hat{\Phi}
+h_{\nu}\hat{l}\hat{H}_u\hat{\nu}^c+h\hat{\Phi}\hat{S}\hat{\nu}^c+\hc,
\ee
which conserves both total lepton number and R-parity (hat denotes a 
superfield). The new superfields
$(\hat{\Phi},\hat{\nu}_i^c,\hat{S}_i)$ are $SU(2)\otimes U(1)$ -singlets 
and carry conserved lepton numbers $(0,-1,-1)$ respectively.

In this paper we shall focus on the $h_e\hat{l}\hat{H}_d\hat{e}^c$ term. 
Using standard methods \cite{haberkane}, one can show that the following 
terms are present in the Lagrangean:
\be{kytkenta}
{\cal L}_R=\overline{\mato{H}}[h_{\nu ij}\mato{\nu}^*_{Li} R+
h_{ekj}\mato{\nu}^*_{Rk} L]e_j,
\ee
where 
\be{hmato}
\mato{H}={\mato{H}_u^+\choose \mato{H}_d^-}
\ee
and $R$, $L$ are the chirality projection operators. 

The Higgsino current associated with charged higgsinos can be 
written as \cite{CQRVW}
\be{hcurrent}
J^{\mu}_{{\mato{H}}}=\overline{{\mato{H}}}\gamma^\mu{\mato{H}}.
\ee
The Higgsino current (\ref{hcurrent}) is associated with a triangle diagram
where Higgsinos interact with electron-like leptons, similar 
to the diagram in ref. \cite{CQRVW}.
The CP-violating source in the diffusion equation is in this case 
created by interactions arising from the term
\be{kytke2}
{\cal L}_R=h_{eij}\overline{{\mato{H}}}v^*_{i} L e_j,
\ee
where $v_i\equiv <\mato{\nu}_{Ri}>$. It is worth noting that a non-vanishing
source requires only that one of the sneutrino fields acquires a vev.
This is due to a complex vev so that two degrees of freedom are present.
In a spontaneously R-parity breaking model the global B-L symmetry 
is spontaneously broken by assuming non-zero vev's for
$v_R\equiv <\mato{\nu}_{R3}>$, $v_S\equiv <\mato{S}_3>$ \cite{masiero}.
In general we may also choose a non-zero value for $v_L\equiv
<\mato{\nu}_{L3}>$. However, spontaneous R-parity breaking generates a 
majoron that are produced in a stellar environment and to suppress
the stellar energy loss one must require
${v_L^2\over v_R}m_W\lsim 10^{-7}$ \cite{masiero}. This is easily
achieved for $v_R=\cO(1\ {\rm TeV})$ provided that $v_L\leq \cO(100\
{\rm MeV})$. For generality we consider here the case where all left- 
and right handed sneutrinos have a non-zero vev.  
In numerical estimations we make the usual assumption that
only $v_{R3}$ is large.

If one chooses interactions similar to ref. \cite{CQRVW} where both 
left- and right-handed electrons are included the
amount of CP-violation is proportional to $\mato{\nu}_R
\mato{\nu}_L$ -term. However, if we consider only interactions like
(\ref{kytke2}), the total CP-violation has a factor $\mato{\nu}_R^2$,
which, assuming that $\abs{v_R}\gg\abs{v_L}$, can give a much larger 
contribution. 

The scenario considered here sets some limitation to the hierarchy of the
electroweak phase transition, namely that the higgsino and sneutrino field(s)
acquire their respective vev's simultaneously.  Otherwise sphaleron transitions
wash out the previously created baryon asymmetry.

The CP-violating source may now be computed using CTP-formalism as  
described in ref. \cite{riotto}.
The interactions of the Higgsinos lead to a contribution to the self-energy
of the form 
\be{scp}
\Sigma_{CP}^<(x,y)=g_{CP}^L(x,y)R
G^{0,<}_{\mato{\nu}}(x,y)L+g^L_{CP}(x,y)LG^{0,<}_{\mato{\nu}}(x,y)R,
\ee
where 
\be{gs}
g^L_{CP}=h^*_{eij}v_{i}(x)h_{ekj}v^*_{k}(y).
\ee
Similarly for the other component $\Sigma_{CP}^>$.
High temperature corrections change the dispersion relation of
charginos and neutralinos \cite{weldon}. The spectral function of
Higgsinos may, in the approximation $\Gamma_{\mato{H}}\ll m_{\mato{H}}$, be 
written as \cite{riotto} \be{spectrl}
\rho_{\mato{H}}({\bf k},k^0)=i(k\slsh +m_{\mato{H}})
\Big[\Big((k^0+i\epsilon+i\Gamma_{\mato{H}})^2-\om_{\mato{H}}^2(k)\Big)^{-1}-
\Big((k^0-i\epsilon-i\Gamma_{\mato{H}})^2-\om_{\mato{H}}^2(k)\Big)^{-1}\Big],
\ee
where $\om_{\mato{H}}^2(k)={\bf k}^2+m^2_{\mato{H}}(T)$. The effective 
squared Higgsino
plasma mass $m_{\mato{H}}^2(T)$ may be approximated by $m_{\mato{H}}^2(T)
\approx |\mu|^2$. Similarly, $\abs{\mu}$ should be replaced by the electron 
plasma mass ${3\over 32}g_2^2T^2$ for $\rho_{e_j}({\bf k},k^0)$.\\
The source term 
\bea{cpsrc}
S_{\mato{H}} & = & -\int d^3r_3\int_{-\infty}^T 
\tr[\Sigma_{CP}^>(X,x_3)
G_{\mato{H}}^<(x_3,X)-G_{\mato{H}}^>(X,x_3)\Sigma_{CP}^<(x_3,X)\non
& + & G_{\mato{H}}^<(X,x_3)\Sigma_{CP}^>(x_3,X)-\Sigma_{CP}^<(X,x_3)
G_{\mato{H}}^>(x_3,X)]
\eea
now contains the following function
\bea{factor1}
f_g & = & g_{CP}^L (X,x_3)-g_{CP}^L(x_3,X)
\non
& = & i\Im[(h^*_{eij}h_{ekj})(v_i(X)v^*_k(x_3)-v_i(x_3)v^*_k(X))],
\eea
which after the Higgs insertion expansion \cite{CQRVW} becomes
\be{factor2}
f_g=i\Im[(h^*_{eij}h_{ekj})(v_i(X)\pa^\mu_Xv^*_k(X)-v^*_k(X)\pa^\mu_X 
v_i(X))].
\ee
The possibility of complex Yukawa coupling constants, $h_e$, deserves
further discussion. The $h_e$ matrix clearly has 18 degrees of freedom
i.e. 9 complex phases are present.
From (\ref{kytke2}) we note that 7 of the complex phases may be set zero
by redefining the sneutrino, higgsino and electron-like lepton fields.
Generally we then have two complex phases left. For simplicity, 
we shall from now on assume that all Yukawa coupling constants are real
i.e. $h_{eij}\in\real,\ i,j=1,2,3$. However, the sneutrino vev's may be 
complex, so that we may write
\be{vevh}
v_i(X)=A_i(X)e^{i\theta_i(X)},
\ee
where $A_i(X),\theta_i(X)\in\real$. Function $f_g$ can now be written as
\be{factor3}
f_g=i(h_{eij}h_{ekj})\Big[\sin(\theta_i-\theta_k)[A_iA_k'-A_kA_i']-
\cos(\theta_i-\theta_k)A_iA_k(\theta_i+\theta_k)'\Big].
\ee
The first term is of the form presented in \cite{CQRVW} with an 
extra factor of $\sin(\theta_i-\theta_k)$. The second term is of a new form,
which exists due to the non-vanishing complex phases of the sneutrino vev's. 

We can now write the expression for the CP-violating source,
\be{sx}
S_{\mato{H}}=2i(h_{eij}h_{ekj})
\Big[\sin(\theta_i-\theta_k)[A_iA_k'-A_kA_i']-
\cos(\theta_i-\theta_k)A_iA_k(\theta_i+\theta_k)'\Big]
{\cal I}^{e_j}_{\mato{H}},
\ee
where ${\cal I}^{e_j}_{\mato{H}}$ is slightly modified from the result 
in \cite{riotto},
\bea{iwh}
{\cal I}^{e_j}_{\mato{H}} & = & \int^\infty_0 dk 
{k^2(\om_{ej}\om_{\mato{H}}-k^2) 
\over 2\pi^2 \om_{ej}\om_{\mato{H}}}\Big[
\Big(1-2\Re(n_{e_j})\Big)F(\om_{\mato{H}},\Gamma_{\mato{H}},\om_{e_j},
\Gamma_{e_j})\non
& + &  \Big(1-2\Re(n_{\mato{H}})\Big)F(\om_{e_j},\Gamma_{e_j},\om_{\mato{H}},
\Gamma_{\mato{H}})+
2\Big(\Im(n_{\mato{H}})+\Im(n_{e_j})\Big)\non
& & G(\om_{\mato{H}},\Gamma_{\mato{H}},\om_{e_j},\Gamma_{e_j})\Big]
\eea
where $n_{\mato{H}}=[\exp\Big(\om_{{\mato{H}}(e_j)}/T+i\Gamma_{{\mato{H}}(e_j)}\Big)
+1]$ and 
\bea{funcs}
F(a,b,c,d) 
& = & {1\over 2}[(a+c)^2+(b+d)^2]^{-1}\sin[2\arctan{a+c\over b+d}]\non
& + & {1\over 2}[(a-c)^2+(b+d)^2]^{-1}\sin[2\arctan{a-c\over b+d}],\non
G(a,b,c,d) 
& = & -{1\over 2}[(a+c)^2+(b+d)^2]^{-1}\cos[2\arctan{a+c\over b+d}]\non
& - & {1\over 2}[(a-c)^2+(b+d)^2]^{-1}\cos[2\arctan{a-c\over b+d}].
\eea
The damping rate of Higgsinos and electron-like leptons is dominated
by weak interactions so that we may take $\Gamma_{\mato{H}}\approx\Gamma_{e_j}\sim
0.05$ T \cite{CQRVW}. 

We shall now review the basic ingredients of electroweak baryogenesis as
presented in ref. \cite{CQRVW}.
In the paper the baryon to entropy ratio was shown to be
\be{bno}
{n_B\over s} = - g(k_i) {{\cal A} \bar D \Gamma_{ws}\over v_w^2 s}, 
\ee
where $g(k_i)$ is a numerical coefficient depending on the degrees of freedom,
$\bar D$ the effective diffusion rate, 
$\Gamma_{ws} = 6 \kappa \alpha_w^4 T$ ($\kappa = 1$) \cite{AK} the weak 
sphaleron rate and $v_w$ the velocity of the bubble wall.
The entropy density $s$ is given by 
\be{entropy}
s = {2 \pi^2 g_{*s} T^3 \over 45},
\ee
where $g_{*s}$ is the effective number of the relativistic degrees of freedom.
${\cal A}$  was shown to be a weighted integral over the CP-violating source
$\tilde \gamma(u) = v_w f(k_i) \partial_u J^0(u)$:
\be{A}
{\cal A} = {1\over \bar D \lambda_+} \int_0^\infty du \tilde \gamma (u) 
e^{-\lambda_+ u},
\ee
where $\lambda_+ = (v_w + \sqrt{v_w^2 + 4 \tilde \Gamma \bar D})/(2\bar D)$ and
the wall was defined to begin at $u = 0$, where $u$ denotes
the co-moving 
coordinate $u = z + v_w t$ and the wall is assumed to move in the direction of
the positive z-axis. ($f(k_i)$ is a 
coefficient depending on the number of degrees of freedom present in thermal 
path and related to the definition of the effective source \cite{CQRVW,Nelson}.)
In the notation of the current paper and after performing a partial
integration (assuming the CP-source vanishes at $u=0$ and 
$u\rightarrow\infty$)
the coefficient ${\cal A}$ is related to the shape of the 
bubble wall through
\bea{integral1}
{\cal A}\propto I_1 & \equiv & \int_0^{\infty} du 
S_{\mato{H}}(u)
e^{- \lambda_+ u}\non
& = & 2i (h_{eij}h_{ekj}) \Big(\int_0^\infty du
\Big[\sin(\theta_i-\theta_k)[A_iA_k'-A_kA_i']\non
& - & 
\cos(\theta_i-\theta_k)A_iA_k(\theta_i+\theta_k)'\Big] e^{-\lam_+u}\Big)
{\cal I}^{e_j}_{\mato{H}}.
\eea
In comparison, in the paper by Carena \etal \cite{CQRVW} the coefficient
${\cA}$ is related to the shape of the bubble wall through
\bea{integral2}
{\cal A}\propto I_2 & \equiv & 
\int_0^{\infty} du \dd{}{u}(H_1H_2'-H_2H_1') e^{-\lam_+ u},\non
& \equiv & \int_0^{\infty} du \dd{}{u}(H(u)^2 \dd{\beta(u)}{u})e^{-\lam_+ u},
\eea
where $H=(H_1^2+H_2^2)^{1\over 2}$, $\tan\beta=H_1/H_2$ and $H_i$'s are 
the real parts of the neutral components of the Higgs doublets.
In the paper by Riotto \cite{riotto}, the CP-violating source 
associated with the charginos was considered and in that case
$I_2$ can be written as
\be{integral3}
{\cal A}\propto I_2 =\Im(\mu) 
[\int_0^{\infty} du H(u)^2\dd{\beta(u)}{u} e^{-\lam_+ u}]
[3M_2 g_2^2{\cal I}^{\mato{W}}_{\mato{H}}+M_1 g_1^2{\cal 
I}^{\mato{B}}_{\mato{H}}],
\ee
where $M_i$ are the soft supersymmetry breaking parameters, 
${\cal I}^{\mato{W},\mato{B}}_{\mato{H}}$ are corresponding
integrals for the winos and the bino similar to (\ref{iwh}).

(\ref{integral1}) can be written as (writing out the sums explicitly)
\bea{int1.2}
I_1 & = & 4i\sum_j \Big[-\sum_i h_{ij}^2\int^\infty_0A_i^2\theta_i'
e^{-\lam_+ u}du\non
& + & \sum_{k<i}h_{ij}h_{kj}\int^\infty_0[\sin(\theta_i-\theta_k)
(A_iA_k'-A_kA_i')\non
& - & \cos(\theta_i-\theta_k)A_iA_k(\theta_i+\theta_k)']
e^{-\lam_+u}du\Big]{\cal I}^{e_j}_{\mato{H}}\non
& \equiv & 4i\sum_j({\cal B}_1+{\cal B}_2+{\cal B}_3){\cal 
I}^{e_j}_{\mato{H}}. \eea
In \cite{CQRVW} the wall shape was assumed to take a simple sinusoidal form 
where the field $H(u)$ can be given by
\be{apwallH}
H(u) = {v \over 2}\left [1 - \cos\left ({u\pi\over L_w} \right )\right ]
[\theta(u) - \theta(u - L_w)] + v \theta(u - L_w)
\ee
and the angle $\beta(u)$ by
\be{apwallbeta}
\beta(u) = {\Delta\beta\over 2} \left [1 - \cos\left ({u\pi\over L_w} 
\right )\right ][\theta(u) - \theta(u - L_w)] + \Delta\beta \theta(u - L_w),
\ee
where $\Delta \beta$ is given by $\Delta\beta = \beta(T_0) - 
\arctan(m_1(T_0)/m_2(T_0))$, calculated at the temperature where the 
curvature
of the one-loop effective potential vanishes at the origin. (It is thus the 
angle between the flat direction and vacuum direction.)
We shall assume that both the modulus and the phase of the sneutrino
vev's have a similar sinusoidal form (\ref{apwallH}).
In this approximation, one can write an analytical expression for
the integral in (\ref{integral3}),
\bea{int2ana}
I_2 & \propto & {\frac{{{\pi }^2}\,\left( 2\,{\lambda_+^4}\,{L_w^4} + 
20\,{\lambda_+^2}\,{L_w^2}\,{{\pi }^2} +
         3\,\left( 11 + 5\,{e^{\lambda_+\,L_w}} \right) \,{{\pi }^4} \right) }
        {4\,{e^{\lambda_+\,L_w}}\,
       \left( {\lambda_+^6}\,{L_w^6} + 14\,{\lambda_+^4}\,{L_w^4}\,{{\pi }^2} +
         49\,{\lambda_+^2}\,{L_w^2}\,{{\pi }^4} + 36\,{{\pi }^6} \right) }}
	\Delta\beta v^2\non
& \equiv & f_1(L_w)\Delta\beta v^2,
\eea
where $L_w$ is the width of the bubble wall and $v^2=<H_1>^2+<H_2>^2$.

Similarly we can now estimate the first term in (\ref{int1.2}),
\be{b1}
{\cal B}_1=-\sum_i h_{eij}^2v_i^2\theta_{f_i} f_1(L_w),
\ee
where $\theta_{f_i}\equiv <\theta_i>$. Here it is assumed that the phases of 
sneutrino fields acquire their vev's sinusoidally from $<\theta_i>=0$ to
$<\theta_i>=\theta_{f_i}$.
${\cal B}_{2,3}$ cannot be
solved analytically. However, before numerical analysis, some special cases 
are worth a further study.
Let us define new phase variables, $\psi_{ik}(X)\equiv\theta_i(X)
-\theta_k(X)$
and $\phi(X)_{ik}\equiv\theta_i(X)+\theta_k(X)$. Now, if 
$\psi_{ik}(X)=\psi_{ik}$, ${\cal B}_{1,3}$ clearly vanish and we are
left with an expression similar to (\ref{int2ana}) with an additional 
$\sin\psi_{ik}$ factor,
\be{erik2}
I_1^+=+4i\sum_j\sum_{k<i}h_{eij}h_{ekj}f_1(L_w)\sin\psi_{ik}(v_i^2+v_k^2)\Delta
\beta_{ik}{\cal I}^{e_j}_{\mato{H}},
\ee
where $\Delta\beta_{ik}$ is the angle between the flat and the vacuum 
direction at the origin.
If, on the other hand, $\psi_{ik}(X)=0$ i.e. all sneutrino fields acquire
the vev of their phases in uniform, we are left with
\be{erik1}
I_1^-=-4i \sum_j\Big[\sum_i h_{eij}^2 v_i^2 \theta_f f_1(L_w)+
2\sum_{k<i}h_{eij}h_{ekj}v_iv_k\theta_ff_1(L_w)\Big]{\cal 
I}^{e_j}_{\mato{H}}, \ee
where $\theta_f\equiv<\theta_i>$ (all phases have the same vev).
Clearly since $\cB_2$ is the only positive term and $\sin\psi_{ik}\leq 1$,
$I_1^+$ gives the maximum value of $I_1$. Similarly $I_1^-$ gives the 
minimum value. However, the overall sign of $I_1$ is not significant
so that we are only concerned with the absolute value of $I_1$. Since
$I_1^+$ and $I_1^-$ have opposite signs, one of them give the maximum
value of $\abs{I_1}$ depending on the choice of parameters. Generally
$I_1^-$ is greater in magnitude than $I_1^+$ due to the $\Delta\beta_{ik}$ 
factor in $I_1^+$. 

To compare the magnitude of the CP-violating source considered in the
present paper with previous results, numerical estimates are in order. In 
choosing parameters we 
shall follow closely ref. \cite{CQRVW}. The bubble wall width is chosen 
to be $L=25/T$ (our results are quite insensitive to the exact choice of the bubble
wall width). 

First we shall consider (\ref{integral3}). We choose $M_2\approx\abs{\mu}
\approx M_1\approx T\approx 100$ GeV and ${\cal 
I}^{\mato{W}}_{\mato{H}}\approx
{\cal I}^{\mato{B}}_{\mato{H}}$ to give us an estimate of the effect
of the CP-violating source. With these values $\sin\theta_\mu$ is roughly
of the order $0.1$ \cite{CQRVW} and $M_2\abs{\mu}{\cal 
I}^{\mato{W}}_{\mato{H}}$ can numerically be shown to have a value
of about $10$. Putting all these values into (\ref{integral3}), we
get an estimate $I_2\approx 0.35 v^2\Delta\beta$. Estimating
$v=246$ GeV and $\Delta\beta\approx 0.015$ \cite{CQRVW} we finally
arrive at an estimate $I_2\sim 300$.
Recent calculations using a more realistic bubble wall shape have lead to an 
additional suppression factor of $\sim 0.3$ \cite{multamaki}.
This effect is ignored here since the analytical approximation 
is used in both cases.

\begin{figure}
\leavevmode
\centering
\vspace*{80mm} 
\includegraphics{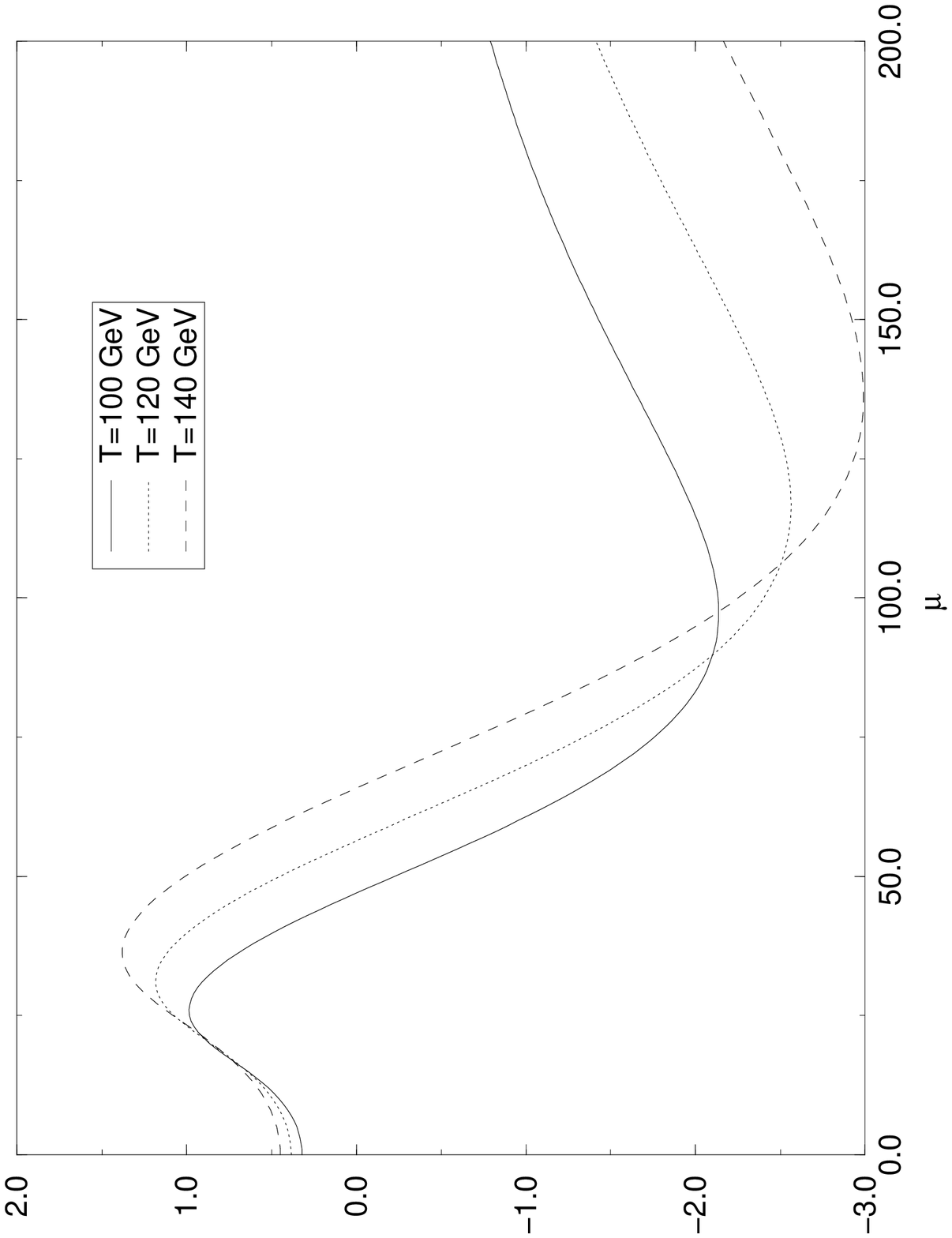}
\caption{Values of ${\cal I}^{e_j}_{\mato{H}}$ at different phase 
transition temperatures.}
\label{kuva} 
\end{figure}

Similarly we can now estimate (\ref{int1.2}). Choosing $L_w=25/T$ we first
find that $f_1(L_w)\approx 0.25$ for $T\sim 100$ GeV. 
Values of ${\cal I}^{e_j}_{\mato{H}}$ are shown in fig. \ref{kuva}.
The values are
very indifferent to the lepton type considered so that
we may very accurately take ${\cal I}^{e_j}_{\mato{H}}=
{\cal I}^{\tau}_{\mato{H}}\equiv{\cal I}_{\mato{H}}$. As from fig. 1
can be seen, if we choose $\mu\approx T\approx 100$ GeV, ${\cal 
I}_{\mato{H}}$ is about $-2$. From the figure it is clear that there 
exists a resonance at value $\mu\approx T$. 
However, this resonance is not as pronounced as in ref. \cite{CQRVW}.

Numerical examinations show that 
the integral in ${\cal B}_2$ is
approximately linear in $\psi_{ik}$ and we may estimate it with
$\sim 0.18 \psi_{ik}$ (no significant $T$ dependence). Furthermore 
the integral in ${\cal B}_3$ is approximately constant and we may take
it to be $\sim 0.05$. Substituting these values into (\ref{int1.2}), we 
then have 
\be{numint1.2}
I_1\approx -i8\sum_j\Big[-0.25\sum_i h_{eij}^2v_i^2\theta_{f_i}+0.18
\sum_{k<i}h_{eij}h_{ekj}(v_i^2+v_k^2)\Delta\beta_{ik}\psi_{ik}
-0.05\sum_{k<i}v_iv_k\phi_{ik}\Big].
\ee
Here we have several unknown parameters. To estimate the
numerical value of (\ref{numint1.2}), let us first assume
that $v_3\gg v_1,v_2$ and $\theta_{f_1}=\theta_{f_2}\approx 0$ i.e.
only the tau-sneutrino (we may have as well chosen any other generation)
has a significantly large vev. In this approximation \be{numint1.3}
I_1\approx -i 8 v_3^2\theta_{f_3}\Big[-0.25\sum_jh^2_{e3j}+
0.18 (\Delta\beta_{31}\sum_jh_{e3j}h_{e1j}+\Delta\beta_{32}
\sum_jh_{e3j}h_{e2j})\Big].
\ee
If we now choose $v_3=1000$ GeV, 
estimate $h_{eij}\sim 10^{-1}$ and take a conservative
estimate for $\Delta\beta_{ij}\sim0.1$,  we get an estimate
$I_1\approx -50000\ \theta_{f_3}$,
which is significantly greater than the 
previous estimate provided that $\theta_{f_3}\gsim 0.01$. 
The values of the
different contributions are in this case $I_1^+\sim 10000\ \sin
\theta_{f_3}$ and $I_1^-\sim -60000\ \theta_{f_3}$.
Even if $h_{eij}\sim\cO(10^{-2})$ and $\theta_{f_3}\sim\cO(1)$,
$I_1$ is still large enough to explain the origin of the baryon asymmetry
of the universe (even when accounting for the extra suppression factor 
due to a more realistic bubble wall profile).

We can now easily estimate the possible baryon to entropy ratio 
resulting from the spontaneously R-parity violating model.
In \cite{CQRVW} it was estimated that with the same choice of parameters used
in this paper, ${n_B\over s}\sim 10^{-11}$. Since $I_1/I_2$ can easily
be $\sim 10^2$ the baryon asymmetry created within the context
of the scenario considered here can be as high as $\sim 10^{-9}$. So the required
value, ${n_B\over s}\sim 4\times 10^{-11}$, is accessible with a non-
restrictive choice of parameters.

In the present paper we have estimated the contribution to CP-violation in
the bubble wall at the electroweak phase transition arising from the terms
present in minimally supersymmetric spontaneously R-parity violating
models. It has been shown that the amount of CP-violation due to the
higgsino-lepton -interaction can quite easily be larger than 
the contribution
due to the higgsino-gaugino -interaction by a factor of $10^2$. 
This contribution is due to the possible existence of a large
(complex) sneutrino vev and a Yukawa coupling constant of order 
$\cO(10^{-1})$. It is notable that this result is
not dependent on the soft supersymmetry breaking parameters $M_{1,2}$ and
furthermore allows freedom in the choice of the value of $\mu$-parameter
which may also be chosen real. Also the phase transition temperature
dependence is quite weak so it can be concluded that this mechanism allows
for the creation of baryon asymmetry at the electroweak phase transition
with quite a flexible choice of parameters. However, it should also be
noted that there are several sources of uncertainty present due to the
approximate nature of the chosen bubble wall profile and how higher
corrections to the CP-violating source behave. Furthermore, the weak
sphaleron rate is under discussion \cite{AJ} and changes on that may
significantly change the results on electroweak baryogenesis. \newpage

\newpage
\end{document}